\documentclass[10pt,journal,compsoc]{IEEEtran}

\usepackage{graphicx}
\usepackage{cite}   
\usepackage{amsmath}
\usepackage{amssymb,amsfonts}
\usepackage{algorithmic}
\usepackage{textcomp}
\usepackage{xcolor}
\usepackage{booktabs} 
\usepackage{epstopdf}
\usepackage{hyperref}
\usepackage{subfig}
\usepackage{bm}

\usepackage{algorithmic}
\usepackage{algorithm}

\usepackage{textcomp}
\graphicspath{{figures/}} 
\newcommand{\revise}{\textcolor{black}}

\usepackage{float}
\def\BibTeX{{\rm B\kern-.05em{\sc i\kern-.025em b}\kern-.08em
    T\kern-.1667em\lower.7ex\hbox{E}\kern-.125emX}}

\ifCLASSINFOpdf

\else

\fi

\begin{document}

    \title{SDR-Empowered Environment Sensing Design and Experimental Validation Using OTFS-ISAC Signals}

\author{
  \IEEEauthorblockN{Jun Wu*},
   \IEEEauthorblockN{Yuye Shi*},
  \IEEEauthorblockN{Weijie Yuan, \IEEEmembership{Senior Member, IEEE}},
    \IEEEauthorblockN{Qingqing Cheng, \IEEEmembership{Member, IEEE}},
   \IEEEauthorblockN{Buyi Li},
  \IEEEauthorblockN{Xinyuan Wei}
\thanks{ This work is supported in part by National Natural Science Foundation of China under Grant 62471208, in part by Guangdong Provincial Natural Science Foundation under Grant 2024A151510098, in part by Shenzhen Science and Technology Program under Grant JCYJ20240813094627037, and in part by ARC DECRA Scheme (DE250101065). An earlier version of this paper was
 presented in part at 2024 IEEE/CIC International Conference on Communications in China (ICCC Workshops), Hangzhou, China\cite{10693814}. (* Co-first authors, corresponding author: Weijie Yuan)

\textbullet~ J. Wu, Y. Shi, B. Li, X. Wei, and W. Yuan are with the School of Automation and Intelligent Manufacturing, Southern University of Science and Technology, Shenzhen, 518055, China (e-mail: \{wuj2021; shiyy2023;   liby2022; weixy2022\}@mail.sustech.edu.cn and yuanwj@sustech.edu.cn).

 \textbullet~ Q. Cheng is with the School of Electrical Engineering and Robotics, Queensland University of Technology, Brisbane, QLD 4000, Australia (e-mail:qingqing.cheng@qut.edu.au).

}
}

\IEEEtitleabstractindextext{%
\begin{abstract}
This paper investigates the system design and experimental validation of integrated sensing and communication (ISAC) for environmental sensing, which is expected to be a critical enabler for next-generation wireless networks. We advocate exploiting orthogonal time frequency space (OTFS) modulation for its inherent sparsity and stability in delay-Doppler (DD) domain channels, facilitating a low-overhead environment sensing design. Moreover, a comprehensive environmental sensing framework is developed, encompassing DD domain channel estimation, target localization, and experimental validation. In particular, we first explore the OTFS channel estimation in the presence of fractional delay and Doppler shifts. Given the estimated parameters, we propose a three-ellipse positioning algorithm to localize the target's position, followed by determining the mobile transmitter's velocity. Additionally, to evaluate the performance of our proposed design, we conduct extensive simulations and experiments using a software-defined radio (SDR)-based platform with universal software radio peripheral (USRP). The experimental validations demonstrate that our proposed approach outperforms the benchmarks in terms of localization accuracy and velocity estimation, confirming its effectiveness in practical environmental sensing applications. 



\end{abstract}

}
\maketitle

\IEEEpeerreviewmaketitle

\section{Introduction}

The next-generation wireless communication networks are anticipated to offer simultaneous high-quality connectivity and precise sensing services, enabling emerging applications, e.g., autonomous driving, low-altitude economy (LAE), and indoor localization \cite{LiuISAC,10663788}. Traditionally, radar sensing and wireless communication have been implemented as separate systems to perform sensing and communication tasks independently, typically using orthogonal frequency bands to avoid mutual interference. However, this frequency segregation fails to utilize spectral resources effectively, leading to inefficient spectrum usage and making it unsustainable for future networks \cite{9557830}. Moreover, both communication and sensing are now seeking additional spectrum resources to meet the rising demands for gigabit-level communication speeds and centimeter-level sensing accuracy. 
Given the scarcity of available spectrum, there is growing interest in spectrum sharing between communication and radar systems \cite{10409528}, facilitated by their similar signal processing techniques and radio frequency (RF) components \cite{9557830,10349836}. While this technology enables access to a broader range of frequency bands, it may also introduce substantial mutual interference \cite{LiuRadar}. To tackle this problem, integrated sensing and communication (ISAC) technology has emerged as a promising solution, which leverages a unified waveform dedicatedly designed to support simultaneous communication and sensing functionalities \cite{10158322,10591333,10418473,10168298}. This integration not only improves spectral efficiency but also provides a pathway for the seamless combination of high-speed data transmission with precise environmental sensing \cite{KimISACsurvey}. 

To realize the promise of ISAC, one of the main challenges is to identify an appropriate waveform that can effectively meet
the requirements of sensing and communications \cite{10107978}.
Modern communication systems predominantly employ orthogonal frequency division multiplexing (OFDM) waveforms for their advantages, including resilience to multipath fading, robust frame synchronization, and low-complexity data detection \cite{YuanVariational}. Meanwhile, OFDM demonstrates commendable performance in radar sensing applications, making it a favorable choice for target detection \cite{5599316}. This capability positions OFDM as a promising candidate for advancing ISAC technology. Existing studies have extensively investigated OFDM-based ISAC systems \cite{7855671,5776640}, employing OFDM waveforms in both bistatic communication and monostatic radar applications \cite{7855671}. Despite the progress achieved, significant challenges remain in OFDM-ISAC systems, such as high peak-to-average power ratio (PAPR) and susceptibility to fast time-varying channels. For instance, in high-mobility environments such as aircraft or high-speed train communications, where speeds typically exceed $100$ kilometers per hour, significant Doppler shifts may occur. These shifts potentially introduce substantial inter-carrier interference (ICI), destroying the orthogonality of OFDM subcarriers and degrading system performance. Against this background, several modulation schemes have been developed, including orthogonal time frequency space (OTFS) \cite{8424569}, generalized frequency division multiplexing (GFDM) \cite{6871292}, orthogonal delay-Doppler division multiplexing (ODDM) \cite{10144350,10516277}, and universal filtered multicarrier (UFMC)\cite{6824990}. Among them, OTFS has garnered paramount attention due to its robustness to delay and Doppler shifts in high-mobility environments \cite{Data-AidedChannelEstimation}. 

OTFS modulates data symbols in the delay-Doppler (DD) domain, rather than the traditional time-frequency (TF) domain in OFDM, offering enhanced resilience to fast time-varying channels and superior error performance \cite{9557830}. This advancement is credited to OTFS's unique modulation and coding strategies, which effectively mitigate the impact of channel variations on the transmitted signal. As substantiated in \cite{9404861}, the set of DD domain basis functions can be meticulously engineered to address the effects of time-selective and frequency-selective fading. Furthermore, the OTFS modulation technique demonstrates a lower peak-to-average power ratio (PAPR) than OFDM, thereby allowing the deployment of high-performance nonlinear power amplifiers.

Inspired by the attractive benefits of OTFS, considerable research effort has been devoted to OTFS-based ISAC systems. Specifically, the study in \cite{9557830} investigated an OTFS-based vehicular communication system where the ISAC technique was employed to support both OTFS uplink and downlink transmissions. In \cite{wu2024low}, the authors proposed a low-complexity minimum bit error ratio (BER) precoder in the DD domain, aiming to enhance the OTFS transmission reliability while guaranteeing the sensing Cramer-Rao lower bound (CRLB). Moreover, \cite{wu2024low} concluded that the DD domain
 channel parameters naturally align with the physical characteristics of the radio propagation environment, enabling OTFS as a promising candidate for fulfilling environment sensing requirements. To this end, the authors in \cite{10151792} developed a hybrid digital-analog architecture to perform environment sensing by leveraging the corresponding Doppler shift and delay estimations. Building on this foundation, the work of \cite{10714398} jointly investigated the target state estimation and transmitter localization, where a weighted least squares (WLS) problem was formulated.
\revise{The simulation results  demonstrated that the proposed OTFS-ISAC framework is effective for environment sensing when leveraging previously estimated target states. Nevertheless, it is important to note that most existing OTFS-based environment sensing studies simplify the system design by assuming integer-tap delay and Doppler shifts. Such an assumption, however, may not hold in practical scenarios with limited time-frequency resources, where the delay and Doppler resolutions are inherently constrained. Consequently, the design of OTFS-ISAC systems that explicitly account for fractional delay and Doppler shifts remains an open and challenging problem. Furthermore, existing works are predominantly validated through simulations, with experimental evaluations in real-world environments still largely absent from the literature.}

Motivated by the above, in this paper, we investigate the environment sensing problem based on OTFS-ISAC techniques, where a framework encompassing system design and experimental validations is established. We adopt an off-grid DD channel estimation method to determine the fractional delay and Doppler, followed by developing a three-ellipse positioning algorithm to localize the target. Furthermore, we employ a software-defined radio (SDR) platform with the universal software radio peripheral (USRP) to evaluate the effectiveness of our proposed design experimentally.
To the best of our knowledge, this is the first work that jointly develops the environment sensing design and experimental validation via OTFS signals. The main contributions of this paper are summarized as follows:
\begin{itemize}
    \item We propose to exploit OTFS signals to perform environment sensing, in contrast to conventional OFDM waveforms. To extract the DD channel parameters in terms of delay and Doppler shift, we investigate a pilot-based off-grid channel estimation algorithm, in which the integer components are first determined, followed by estimating the fractional parts. 
    \revise{
  \item We develop a novel three-ellipse positioning framework to accurately localize target positions. The target localization problem is reformulated as a nonlinear matrix equation, which is then solved through a double weighted least squares (WLS) optimization approach to obtain a robust suboptimal solution. Building upon the estimated target positions, the transmitter's velocity is subsequently derived using an efficient least squares (LS) estimation method.
    \item  To validate the proposed framework under realistic conditions, we implement a SDR platform utilizing USRP devices for practical channel state information acquisition. Extensive simulations and experiments are conducted to evaluate the effectiveness of our proposed design based on the collected data. The proposed scheme achieves superior localization and velocity estimation performance compared to the benchmarks, demonstrating its potential for environment sensing in real-world applications.}
\end{itemize}
\begin{figure*}[tbp]
  \includegraphics[width =\textwidth]{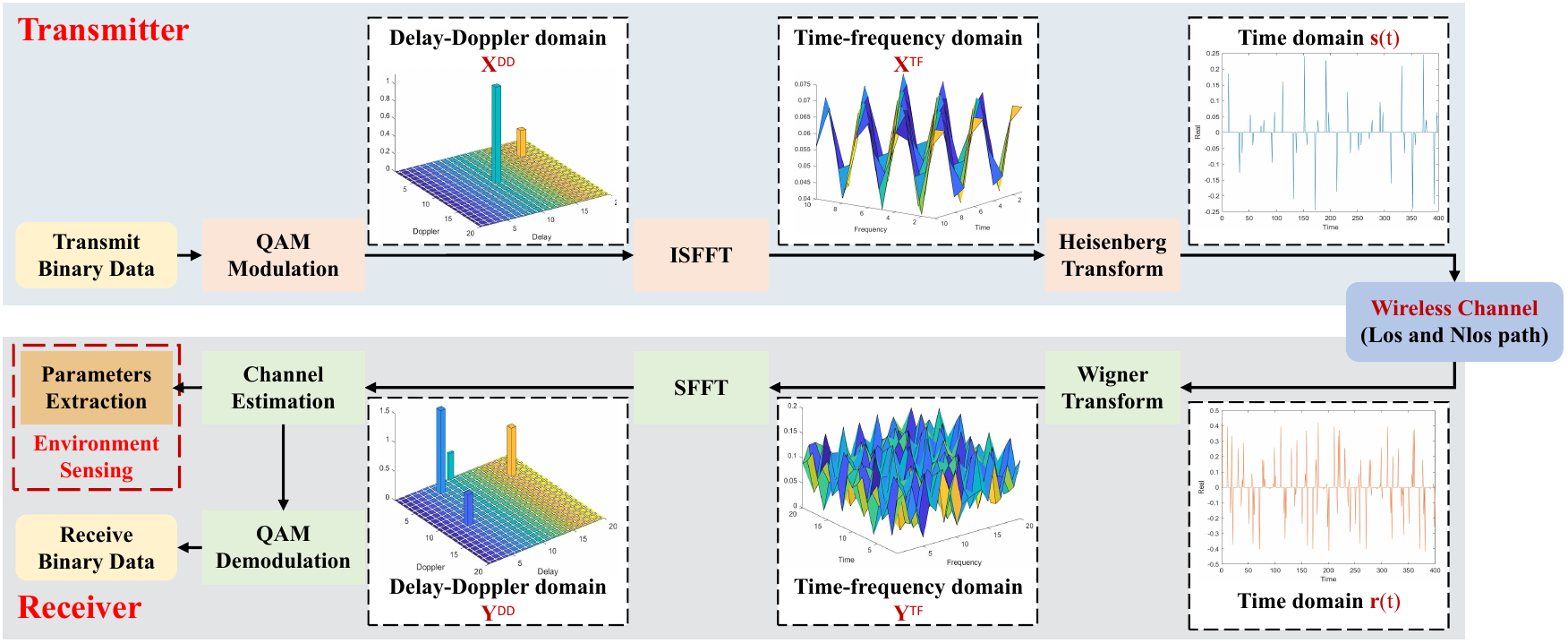}
  \centering
  \caption{Schematic diagram of the OTFS modulation/demodulation.}
  \label{fig:OTFSSystem}
\end{figure*}
The remainder of this paper is organized as follows: Section II introduces the OTFS communication and environment sensing model. The pilot-based off-grid channel estimation algorithm is presented in Section III, and Section IV provides details of the proposed environment sensing. Section V presents the simulation and experimental validation results, followed by Section VI concluding this paper.

\textit{Notations:} Unless otherwise specified, the boldface lowercase letter and boldface capital letter denote the vector and the matrix, respectively. The superscript $(\cdot)^{T}$ indicates the transposition operation.  $|\cdot|$, $|| \cdot||$, and $\mathbb{E}(\cdot)$ refer to the modulus, the $2$-norm and the expectation operation, respectively. We define $\mathbb{C}$, $\mathbb{R}$, and $\mathbb{Z}$ as the set of complex numbers, the real numbers, and the integers, respectively. $\mathbf{A}{[n,m]}$ is the $(n,m)$-th entry of matrix $\mathbf{A}$. $\mathrm{cov}(\mathbf{b})$ stands for the covariance of the vector $\mathbf{b}$. For a time-dependent function $x(t)$, the first-order derivative with respect to time $t$ is expressed as $\dot{x}(t)$. We set $\mathrm{diag}(\mathbf{a})$ and $\mathbf{I}_{M}$ to generate a diagonal matrix with $\mathbf{a}$ being the diagonal elements and to represent the $M$-dimensional identity matrix, respectively.


\section{System Model}
\label{sec2}
We consider a typical OTFS-enabled ISAC system with a mobile transmitter (Tx), a fixed receiver (Rx), and a target to be positioned\footnote{\revise{
We focus on reliable communication leveraging OTFS modulation, integrating sensing functionality into the communication signals. A joint design of communication and sensing tailored for the OTFS framework is left to our future investigation.
}}. Note that both the Tx and Rx are equipped with a USRP platform with a single antenna to transmit/receive OTFS signals for performance validation. In what follows, we first present the OTFS modulation and demodulation, followed by introducing the OTFS-enabled environment sensing model.
\subsection{OTFS Modulation/Demodulation}
At the Tx, the information symbols are first encoded into a two-dimensional (2D) matrix in the DD domain, denoted as $\mathbf{X}^{\mathrm{DD}} \in \mathbb{C}^{M\times N}$, where $M$ and $N$ represent the maximum numbers of delay and Doppler taps, respectively. By applying the inverse symplectic Fourier transform (ISFFT), $\mathbf{X}^{\mathrm{DD}}$ can be converted into the TF domain, say $\mathbf{X}^{\mathrm{TF}}$. In particular, the
entry in the $n$-th row and $ m$-th column of $\mathbf{X}^{\mathrm{TF}}$ is given by
\begin{equation}
\mathbf{X}^{\mathrm{TF}}[n,m]=\frac{1}{\sqrt{MN}}\sum_{k=0}^{N-1} \sum_{l=0}^{M-1} {X}^{\mathrm{DD}}[k,l]e^{-2j\pi (\frac{ml}{M}-\frac{nk}{N})},
\end{equation}
where $k \in \{0,\cdots, N-1 \}$ and $l\in \{0,\cdots, M-1 \}$ are the Doppler and delay taps, respectively. By performing the Heisenberg transform on $\mathbf{X}^{\mathrm{TF}}$, the time signal ${s}(t)$ is then obtained by
\begin{equation}
{s}(t)=\sum_{m=0}^{N-1} \sum_{n=0}^{M-1} \mathbf{X}^{\mathrm{TF}}[n,m]e^{2j\pi m\triangle f(t-nT)}g_{\mathrm{tx}}(t-nT), 
\end{equation}
where $T$, $\triangle f$, and $g_{\mathrm{tx}}(t)$ are the sampling time, frequency intervals, and transmitting pulse shaping filter, respectively. Then, the time domain signal ${s}(t)$ is transmitted passing through time-varying channels. 
Without loss of generality, we jointly take into account the line-of-sight (LoS) path as well as the non-LoS (NLoS) path. Thus, the channel response can be expressed as\cite{10349836}
\begin{equation}
    {h}(\tau ,\upsilon )=\sum_{j=1}^{P} h_j\delta (\tau -\tau_j)\delta (\upsilon -\upsilon_j), \label{chanres}
\end{equation}
where $P$ is the number of resolvable paths and $h_j$ denotes the path gain. $\tau_j=\frac{l_{\tau_j}}{M \Delta f}$ and $\upsilon_j=\frac{k_{\upsilon_j}}{NT}$ represent the delay and the Doppler shifts associated with the $j$-th path, with delay tap and Doppler tap $l_{\tau_j}$ and $k_{\upsilon_j}$, respectively. Note that $l_{\tau_j}$ and $k_{\upsilon_j}$ both include integer and fractional components. The corresponding received signals can be expressed as
\begin{equation}
    {r}(t) = \iint {h}(\tau ,\upsilon ){s}(t-\tau )e^{j2\pi \upsilon (t-\tau)}d\tau d\upsilon+z(t),
\end{equation}
where $z(t)$ is the additive white Gaussian noise (AWGN). ${r}(t)$ is then transformed into a time-frequency signal $\mathbf{Y}^{\mathrm{TF}}[n,m]$ at the Rx by conducting the discrete Wigner transform, which is 
\begin{equation}
     \mathbf{Y}^{\mathrm{TF}}[n,m]=\left[\int g_{\mathrm{rx}}\left(t-\tau\right) {r}(t)e^{-j2\pi\upsilon(t-\tau)}dt\right]\mid_{\tau=nT,\upsilon=m\Delta f},
\end{equation}
where $g_{\mathrm{rx}}(t)$ is the receiving filter. Without loss of generality, we assume that both $g_{\mathrm{tx}}(t)$ and  $g_{\mathrm{rx}}(t)$ are rectangular pulse shapings \cite{wu2024low}.  Subsequently, we employ the symplectic finite Fourier transform (SFFT) to obtain the DD signal $\mathbf{Y}^{\mathrm{DD}}[k,l]$, by
\begin{equation}
\mathbf{Y}^{\mathrm{DD}}[k,l]=\frac{1}{\sqrt{MN}}\sum_{n=0}^{N-1}\sum_{m=0}^{M-1} \mathbf{Y}^{\mathrm{TF}}[n,m]e^{2j\pi\left(\frac{ml}{M}-\frac{nk}{N}\right)}+\Tilde{z}[k,l],
\end{equation}
with $\Tilde{z}[k,l]$ being the corresponding AWGN in the DD domain. The process of OTFS modulation and demodulation is illustrated in Fig. \ref{fig:OTFSSystem}.

\begin{figure}[!tpb]
  \includegraphics[width = .9\linewidth]{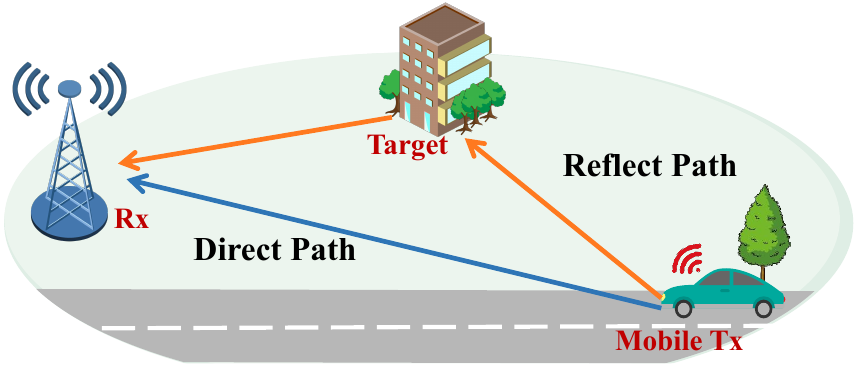}
  \centering
  \caption{The considered environment sensing scenario.}
  \label{fig:Illustration of environment sensing system model}
\end{figure}

\subsection{Environment Sensing Model}
In this section, we investigate the environment sensing model by leveraging the property of OTFS signals. Specifically, the OTFS signal propagates through a multi-path channel, introducing distinct delays due to various propagation distances across paths.
By analyzing the input-output relationship of OTFS signals, it becomes possible to perform channel estimation for extracting accurate delay and Doppler shifts. These components, in turn, provide valuable information for determining the corresponding distance and radial velocity. With the coordinates Tx and Rx, the target's location and the mobile Tx's velocity can be readily determined, enabling effective environment sensing.

We consider a scenario with $P=2$ resolvable paths from the mobile Tx to Rx, i.e., one LoS path and one NLoS path reflected by the target\footnote{\revise{In this work, a simplified two-path model is employed to establish a foundational framework for DD-domain environmental sensing, facilitating experimental validations. However, the proposed environmental sensing approach remains effective even in scenarios involving multi-path propagation. }}, as depicted in Fig. \ref{fig:Illustration of environment sensing system model}. We assume that the Rx's position is fixed at  $\mathbf{t}=[0,0]^\mathrm{T}\in \mathbb{R}^{2\times 1}$, while the time-varying coordinates of Tx at the $i$-th time instant is denoted by $\mathbf{s}_i\in \mathbb{R}^{2\times 1}$. The velocity of the mobile Tx at the $i$-th time instant and the position of the target are denoted by $\dot{\mathbf{s}}_i\in\mathbb{R}^{2\times 1}$ and $\mathbf{p}\in \mathbb{R}^{2\times 1}$, respectively, which are to be estimated.  
As demonstrated in Fig. \ref{fig:Illustration of environment sensing system model}, the distance of the NLoS path is written as
\begin{equation}\label{eqforri}
r_{i}=\left\|\mathbf{p}-\mathbf{s}_i\right\|+\left\|\mathbf{p}\right\|. 
\end{equation}
It is noteworthy that the distances from the Tx to the target and Rx both vary due to the movement of Tx. The time of arrival (ToA) of the NLoS path at $i$-th instant is $T_i$, which also corresponds to the propagation delay of the NLoS path. Then, the NLoS path length can be obtained, expressed as ${r}_i = cT_i$ with $c$ being the light speed. The radial velocity along the direction of the NLoS channel is denoted by $\dot{r}_{i}$, which is 
\begin{equation}\label{eqforridot}
\dot{r}_{i}=\frac{(\mathbf{s}_i-\mathbf{p})^\mathrm{T} \dot{\mathbf{s}}_i}{\left\|\mathbf{s}_i-\mathbf{p}\right\|}.
\end{equation}
Similarly, the LoS path length can be represented by $d_{i}=\left\|\mathbf{s}_i-\mathbf{t}\right\|$, and the radial velocity along the LoS path can be written as 
\begin{equation}\label{eqfordidot}
\dot{d}_{i}=\frac{(\mathbf{s}_i-\mathbf{t})^\mathrm{T} \dot{\mathbf{s}}_i}{d_{i}}.
\end{equation}
It is noted that $\dot{r_i}$ and $\dot{d}_{i}$ are obtained from corresponding Doppler shifts, i.e., ${\dot{r}}_i = -\frac{cv_{r_i}}{f_c}$ and ${\dot{d}}_i = -\frac{cv_{d_i}}{f_c}$, where ${f_c}$ is the carrier frequency. $v_{r_i}$ and $v_{d_i}$ represent the Doppler shifts associated with the NLoS path and LoS path at the $i$-th instant, respectively.

In practice, the variables ${r}_i$, ${\dot{r}}_i$, and ${\dot{d}}_i$ are affected by estimation errors due to uncertainties in the inferred delay and Doppler, such that we have
\begin{equation}\label{eqforerror}
\begin{aligned}
e_i &= r_i - \hat{r}_i, \\
e_{\dot{r}_i} & = \dot{r}_i - \hat{\dot{r}}_i, \\
\text{ and} \ e_{\dot{d}_i} & = \dot{d}_i - \hat{\dot{d}}_i,
\end{aligned}
\end{equation}
where $\hat{r}_i$, $\hat{\dot{r}}_i$ and $\hat{\dot{d_i}}$ are the estimations corresponding to their respective true values. $e_i$, $e_{\dot{r_i}}$ and $e_{\dot{d_i}}$ represent the errors between the estimated and actual values. For ease of mathematical derivation, these errors are assumed to obey Gaussian distributions with zero-mean and variances of $\sigma_i^2$, $\sigma_{\dot{r_i}}^2$, and $\sigma_{\dot{d_i}}^2$, respectively, as commonly adopted in \cite{10714398,li2023otfs}.

\section{OTFS Channel Estimation}
\label{sec3}
\begin{figure} [t!]
	\centering
	\includegraphics[width=0.99\linewidth]{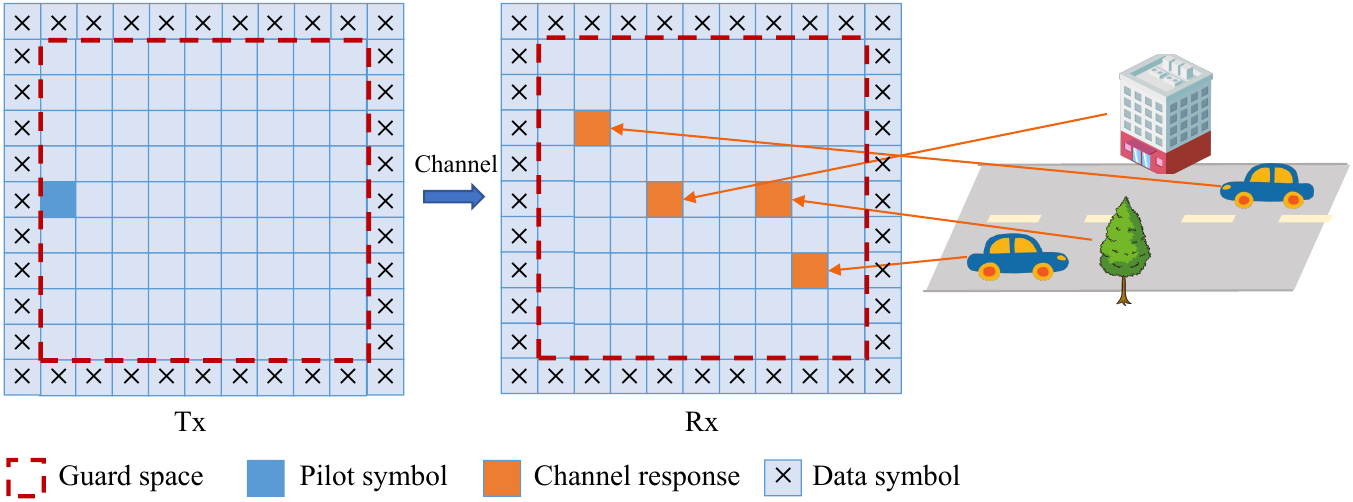}
	\caption{\revise{The pilot symbol placement scheme at the mobile Tx, showing the channel response at the Rx influenced by a multi-path propagation environment with $4$ scatterers.}}
	\label{pilot} 
\end{figure}
As discussed above, accurate delay and Doppler estimations are essential for effective environmental sensing. To this end, we adopt a pilot-based channel estimation method in this section, where the pilot symbol is embedded within the transmitted data symbols and surrounded by a guard space to avoid interference caused by data symbols \cite{WeiXinyuanIEEE}. In particular, the pilot symbol is placed by 
\begin{equation}
\mathbf{X}^{\mathrm{DD}}= \begin{cases}x_p, & k=k_p, l=l_p, \\ 0, & k \in\left[k_p-k_{\max }, k_p+k_{\max }\right] \backslash k_p, \\ & l \in\left[l_p, l_p+l_{\max }\right] \backslash l_p,\\ \text { data symbol, } & \text { otherwise. }\end{cases}.
\end{equation}
The pilot symbol $x_p$ is located at $(k_p, l_p)$. $l_{\max }$ and $k_{\max }$ denote the maximum delay and Doppler taps, respectively. The guard symbols ensure that the pilot remains within the guard space after passing through the DD channel. Fig. \ref{pilot} provides an example of the pilot placement scheme with $4$ scatters in the channel. As illustrated in this figure, $4$ distinct channel responses are contributed by scatters, which can be readily distinguished due to their unique propagation delays and Dopplers. This separation enables effective channel estimation, which, in turn, allows for the extraction of physical parameters of the target, e.g., its distance and radial velocity relative to the Tx, thereby facilitating environmental sensing.

To proceed, we first estimate the delay and Doppler taps.
The channel response with respect to the pilot symbol is expressed as 
\revise{
 \begin{equation}
 \begin{split}
         \mathbf{H}^{\mathrm{DD}}[k,l]=\frac{1}{MN}\sum_{j=1}^{P}h_{j}\sum_{n=0}^{N-1}e^{-j2\pi n(k-k_{\upsilon_{j}})/N}\\\sum_{m=0}^{M-1}e^{j2\pi m(l-l_{\tau_{j}})/M}
         \times e^{-j2\pi\frac{ k_{\upsilon_{j}}l_{\tau_{j}}}{NM}}.
          \end{split}
 \end{equation}
In general, the delay and Doppler shifts cannot be perfectly aligned with the DD domain grid using only integer taps due to the limited resolutions. As a result, the taps $k_{\upsilon_{j}}$ and $l_{\tau_{j}}$ both involve integer and fractional terms, i.e., $k_{\upsilon_{j}}=k_j+\kappa_j$ and $l_{\tau_{j}}=l_j+\iota_j$, where $k_j\in \mathbb{Z}$ and $l_j\in \mathbb{Z}$ are the integer components associated with the $j$-th path, while the fractional parts are represented by $\kappa_j\in[-0.5,0.5]$ and $\iota_j\in[-0.5,0.5]$, respectively.} In such a context, it is intractable to obtain $k_{\upsilon_{j}}$ and $l_{\tau_{j}}$ directly \cite{10118609}. To address this problem, we determine the channel parameters via a two-step method, in which we first derive the integer parts, followed by estimating the fractional components. 
 
\begin{figure}[tbp]
\centering
\includegraphics[width=\linewidth]{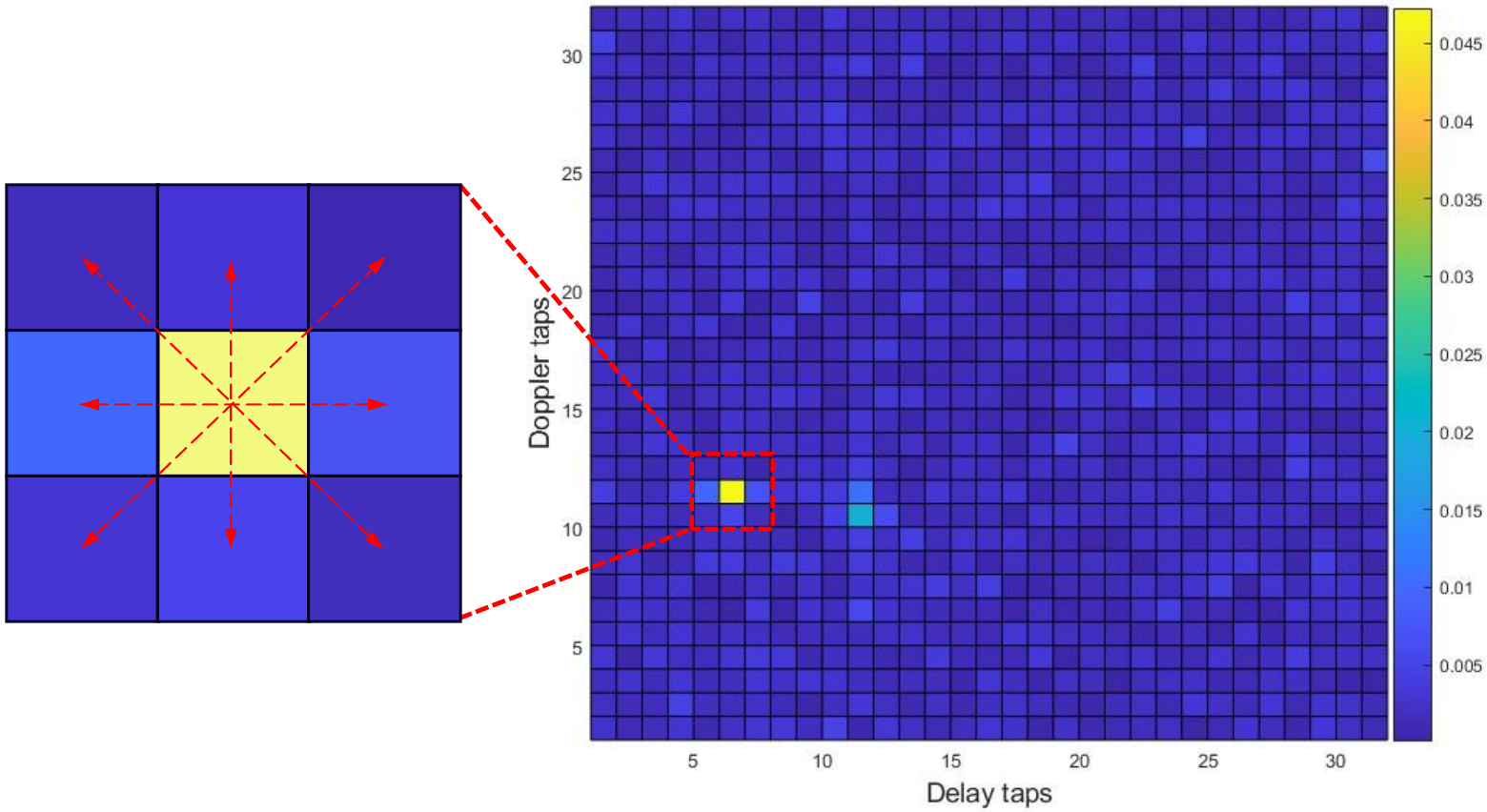}
\caption{The illustration of determining the integer parts by finding the peaks.}
\label{fig:regionmax}
\end{figure}

Intuitively speaking, the integer terms of the delay and Doppler taps can be obtained by identifying the prominent peaks. To achieve this, we set a $3\times 3$ square region centered at the point of interest as a sliding window \cite{10279816}, as shown in Fig. \ref{fig:regionmax}.
A value located at the center of the window is considered a potential peak if it surpasses the values of its all adjacent neighbors. 
The coordinates set of the neighbors, surrounding a center point at $[k,l]$, are designated by $\mathcal{N}_{k,l}$. Note that, noise can cause multiple spurious peaks in practical scenarios, complicating the peak-searching process. To address this, the pilot symbols can be assigned higher energy levels than the data symbols \cite{10118609}. Considering $\mathcal{\tilde{P}}$ qualified potential peaks in total, the largest $P$ values can be regarded as the channel responses of the $P$ paths. Moreover, the coordinates associated with $P$ peak values are the corresponding integer components of delay/Doppler taps.
In a nutshell, the integer part determinations can be mathematically summarized as
\begin{equation}
    \label{equklfinal}
    \begin{cases}
    [k_{1},l_{1}]=\underset {[k,l] \in \mathcal{M}}{\operatorname*{argmax}}\ \mathbf{H}^{\mathrm{DD}}[k,l],\\
    \vdots \\
    [k_{j},l_{j}]=
    \underset {[k,l] \in \mathcal{M} \backslash\{[k_1,l_1],\cdots,[k_{j-1},l_{j-1}]\}}{\operatorname*{argmax}}\mathbf{H}^{\mathrm{DD}}[k,l],\\
    \vdots \\
    [k_{P},l_{P}]=
    \underset {[k,l] \in \mathcal{M} \backslash\{[k_1,l_1],\cdots,[k_{P-1},l_{P-1}]\}}{\operatorname*{argmax}}
    \mathbf{H}^{\mathrm{DD}}[k,l],
    \end{cases}
\end{equation}
where $\mathcal{M}=\{ [k,l] | \mathbf{H}^{\mathrm{DD}}[k,l]>\mathbf{H}^{\mathrm{DD}}[k',l'], [k',l'] \in \mathcal{N}_{k,l} \}$. \revise{
It is worth highlighting that the integer parts of the delay and Doppler taps are determined through a localized search restricted to the pilot guard space, which ensures that the proposed channel estimation scheme remains computationally efficient.
}

Next, the fractional parts $\kappa_j$ and $\iota_j$ of $j$-th path are estimated. Due to space constraints, we take the fractional Doppler tap estimation as a study case and omit fractional delay tap estimation as it follows a similar approach. With the integer taps $l_j$ and $k_j$,  the channel response can be expressed as
\revise{
\begin{equation}\label{eqAkl}
\begin{aligned}
&\mathbf{H}^{\mathrm{DD}}[k,l_{j}] \\& =\frac{h_{j}}{MN}\sum_{n=0}^{N-1}e^{-j2\pi n\frac{k-k_{\kappa_j}}{N}}\sum_{m=0}^{M-1}e^{j2\pi m\frac{(l_j-l_{{\tau_j}})}{M}}e^{-j2\pi\frac{k_{\kappa_j}l_{\tau_j}}{NM}}. 
\end{aligned}
\end{equation}
Notice that $
\sum_{m=0}^{M-1} e^{j2\pi m \frac{\Delta l}{M}} = e^{j\pi \Delta l \left(1-\frac{1}{M}\right)} \frac{\sin(\pi \Delta l)}{\sin\left(\pi \frac{\Delta l}{M}\right)}
$ with $\Delta l=l_j-l_{\tau_j}$. With this property,  the magnitude $|\mathbf{H}^{\mathrm{DD}}[k,l_j]|$ can be further expressed as \cite{5438786}
\begin{equation}\label{magnitude}
    |\mathbf{H}^{\mathrm{DD}}[k,l_j]|=\frac{|h_j|}{MN} \left| \frac{\sin \left(\pi\left(k-k_{\kappa_j}\right)\right)}{\sin \left(\frac{\pi\left(k-k_{\kappa_j}\right)}{N}\right)}\right|\left| \frac{\sin \left(\pi\left(l_j-l_{\tau_j}\right)\right)}{\sin \left(\frac{\pi\left(l_j-l_{\tau_j}\right)}{M}\right)}\right|.
\end{equation}
As stated in \cite{WeiXinyuanIEEE}, the peak is positioned between the two sampling points $\Tilde{k}_1$ and $\Tilde{k}_2$ which are 
\begin{equation}\label{k1k2max}
    \begin{cases}
        \Tilde{k}_1=\underset {k \in \{0, \cdots,  N-1\}}{\operatorname*{argmax}} |\mathbf{H}^{\mathrm{DD}}[k,l_j]|,\\
        \Tilde{k}_2=\underset {k \in \{ \Tilde{k}_{1} + 1, \Tilde{k}_{1} - 1\}}{\operatorname*{argmax}} |\mathbf{H}^{\mathrm{DD}}[k,l_j]|.
    \end{cases}
\end{equation}
Obviously, $\Tilde{k}_1$ is the integer part, i.e., $\Tilde{k}_1=k_j$.  Relying on  \eqref{magnitude}, the ratio of magnitudes at $k=\Tilde{k}_1$ and $k=\Tilde{k}_2$ is derived as
\begin{equation}\label{ratio}
\begin{aligned}
\frac{\left|\mathbf{H}^{\mathrm{DD}}[\Tilde{k}_1,l_j]\right|}{\left|\mathbf{H}^{\mathrm{DD}}[\Tilde{k}_2,l_j]\right|}&=  \left| \frac{\sin \left(\pi\left(\Tilde{k}_{1}-k_{j}-\kappa_{j}\right)\right)}{\sin \left(\frac{\pi\left(\Tilde{k}_{1}-k_{j}-\kappa_{j}\right)}{N}\right)}\right| \\
& \cdot\left| \frac{\sin \left(\pi\left(\Tilde{k}_{2}-k_{j}-\kappa_{j}\right)\right)}{\sin \left(\frac{\pi\left(\Tilde{k}_{2}-k_{j}-\kappa_{j}\right)}{N}\right)}\right|^{-1} \\
& =\left|\frac{\sin \left(-\kappa_{j} \pi\right)}{\sin \left(\frac{\kappa_{j} \pi}{N}\right)}\right| \cdot\left|\frac{\sin \left(\pi\left(\Tilde{k}_{2}-k_{j}-\kappa_{j}\right)\right)}{\sin \left(\frac{\pi\left(\Tilde{k}_{2}-k_{j}-\kappa_{j}\right)}{N}\right)}\right|^{-1}.
\end{aligned}
\end{equation}}
In light of \eqref{k1k2max}, we note the fact that $| \Tilde{k}_{2}-k_j | = 1$, which means 
\begin{align}
  \left|\sin \left(-\kappa_{j} \pi\right)\right|=\left|\sin \left(\pi\left(\Tilde{k}_{2}-k_{j}-\kappa_{j}\right)\right)\right|.  \label{sim}
\end{align} 
Plugging (\ref{sim}) into \eqref{ratio}, the ratio can be further simplified as
\begin{equation}\label{ratioSimple}
\begin{aligned}
\frac{\left|\mathbf{H}^{\mathrm{DD}}[\Tilde{k}_1,l_j]\right|}{\left|\mathbf{H}^{\mathrm{DD}}[\Tilde{k}_2,l_j]\right|}& =\left\lvert\, \frac{\sin \left(\frac{\pi\left(\Tilde{k}_2-k_{j}-\kappa_{j}\right)}{N}\right)}{\sin \left(\frac{-\kappa_{j} \pi}{N}\right)}\right\lvert \approx\left|\frac{\Tilde{k}_2-k_{j}-\kappa_{j}}{-\kappa_{j}}\right|.
\end{aligned}
\end{equation}
Hence, the fractional part of the Doppler tap can be obtained by \cite{shi2021deterministic}
\begin{equation}
    \kappa_j=\frac{(k_j-k_j')\left\lvert \mathbf{H}^{\rm{DD}}[k_j,l_j]\right\rvert}{\left\lvert \mathbf{H}^{\rm{DD}}[k_{j},l_{j}]\right\rvert+\left\lvert \mathbf{H}^{\rm{DD}}[k_{j}',l_{j}]\right\rvert}, \label{dopplercal}
\end{equation}
where $k_j'$ corresponds to $\Tilde{k}_2$.
Similarly, we can also achieve fractional delay tap estimation, by
\begin{equation}
    \iota_j = \frac{(l_{j}-l_{j}')\left\lvert \mathbf{H}^{\rm{DD}}[k_{j},l_{j}]\right\rvert}{\left\lvert \mathbf{H}^{\rm{DD}}[k_{j},l_{j}]\right\rvert+\left\lvert \mathbf{H}^{\rm{DD}}[k_{j},l_{j}']\right\rvert}, \label{delaycal}
\end{equation}
where the definition of \( l_j' \) parallels that of \( k_j' \). Consequently, the taps \( k_{\upsilon_j} \) and \( l_{\tau_j} \) can be determined by aggregating their fractional and integer components, respectively. Then, the delay and Doppler shift are respectively derived by \cite{9927420} 
\begin{equation}\label{delayDopplerget}
    \begin{cases}
        \tau_j=\frac {l_{\tau_j}}{B},\\
        \upsilon_j=\frac {k_{\upsilon_j}B}{MN},
    \end{cases}
\end{equation}
where $B$ is the bandwidth.  The overall algorithm for estimating the delay and Doppler shifts is provided in Algorithm \ref{alg1}.

\begin{figure}[t]
  \begin{minipage}{\linewidth}
  \begin{algorithm}[H]
    \renewcommand{\algorithmicrequire}{\textbf{Input:}}
    \renewcommand{\algorithmicensure}{\textbf{Output:}}
    \caption{ Delay and Doppler shift estimation algorithm} 
    \label{alg1} 
    \begin{algorithmic}[1]
      \REQUIRE $\mathbf{H}^{\mathrm{DD}}\left[k,l\right]$ 
      \STATE Based on \eqref{equklfinal}, obtain the integer parts of $k_{\upsilon_{j}}$ and $l_{\tau_{j}}$, denoted by  $k_{j}$ and $l_{j}$.
      \STATE Based on \eqref{dopplercal} and \eqref{delaycal}, obtain the fractional parts of $k_{\upsilon_{j}}$ and $l_{\tau_{j}}$, denoted by $\kappa_j$ and $\iota_j$.
      \STATE Calculate $l_{\tau_{j}} = l_{j} + \iota_j$.
      \STATE Calculate $k_{\upsilon_{j}} = k_{j} + \kappa_j$.
      \STATE Calculate $\tau_j$ and $\upsilon_j$ according to \eqref{delayDopplerget}.
       \ENSURE Delay $\tau_j$  and Doppler shift $\upsilon_j$.
    \end{algorithmic} 
  \end{algorithm}
  \end{minipage}
\end{figure}

\section{Target Positioning And Velocity Estimation}
\label{sec4}

In this section, we propose a double WLS-based method to obtain the target's position based on the estimated delay and Doppler shifts. Subsequently, the velocity of the mobile Tx is determined via the LS approach incorporating the obtained target location. 

\begin{figure}[t]
	\centering
	\includegraphics[width=0.8\columnwidth]{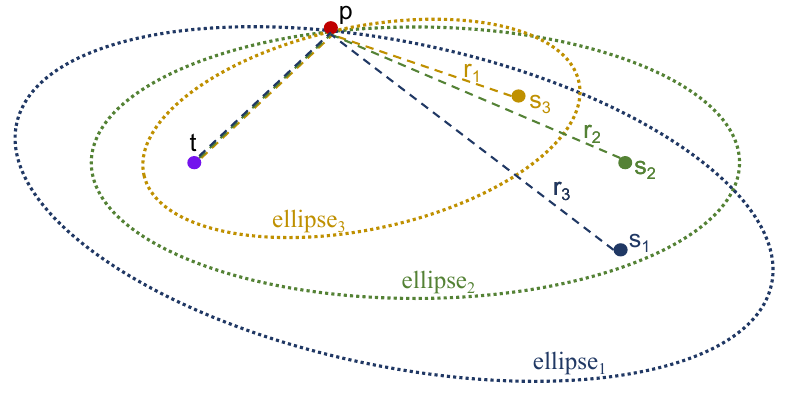}
	\caption{An illustration of ellipse-based positioning approach.}
	\label{EllipseAlgorithm}
\end{figure}

At the $i$-th time instant, the distance of the NLoS path $\hat{r}_i$ can be estimated via \eqref{eqforri}. In light of the principles of time difference of arrival (TDoA) \cite{4667729},  the target is located on an ellipse, whose focal points are the location of the Tx and the Rx, respectively. Since the Tx is mobile, it generates multiple ellipses that all transit the location of the target at different time instants, as illustrated in Fig. \ref{EllipseAlgorithm}. Theoretically, at least three ellipses are required to unambiguously localize the target \cite{10158322}. Each ellipse provides additional positional information that allows for a more precise intersection, ultimately enabling a unique determination of the target's position in a three-dimensional space. \revise{Due to the presence of measurement errors, the multiple ellipses may not intersect at a unique point. To handle this challenge, we reformulate the target localization as a nonlinear matrix equation and employ a double WLS method in what follows, ensuring accurate localization.}

Let us take a further look at \eqref{eqforri}, which can be transformed into the following forms through some simple algebraic derivations, which is
\begin{align}
    \frac{1}{2}(r_i^2 - \|\mathbf{s}_i\|^2) - r_i\|\mathbf{p}\| + \mathbf{s}_i^\mathrm{T}\mathbf{p} = 0.
\end{align}
By incorporating the measurement error \(r_i - \hat{r}_i = e_i\), we then have 
\begin{equation}
\frac{1}{2}(\hat{r}_i^2-\left\|\mathbf{s}_i\right\|^2)-\hat{r}_i\left\|\mathbf{p}\right\|+\mathbf{s}_i^\mathrm{T}\mathbf{p} \approx e_i(\left\|\mathbf{p}\right\|-\hat{r}_i),\label{dew}
\end{equation}
in which the second-order error term \(e_i^2\) is disregarded, enabling the measurement error to be expressed in a linearized form while preserving analytical tractability. By aggregating the observations over \(N_1\ge3\) time instants, \eqref{dew} can be re-expressed into a compact matrix form, as
\begin{equation}
\bm{\alpha} -\mathbf{A}\mathbf{z} \approx \mathbf{B}\mathbf{e},  \label{asd}  
\end{equation}
where
\begin{equation}
\begin{aligned}
\bm{\alpha} &=\left[\begin{array}{c}
\frac{1}{2}(\hat{r}_1^2-\left\|\mathbf{s}_1\right\|^2) \\
\frac{1}{2}(\hat{r}_2^2-\left\|\mathbf{s}_2\right\|^2) \\
\vdots \\
\frac{1}{2}(\hat{r}_{N_1}^2-\left\|\mathbf{s}_{N_1}\right\|^2)
\end{array}\right]\in \mathbb{R}^{N_1\times 1},  \\
\mathbf{A} &=-\left[\begin{array}{ccc}
\mathbf{s}_1^\mathrm{T} & -\hat{r}_1 \\
\mathbf{s}_2^\mathrm{T} & -\hat{r}_2 \\
& \vdots \\
\mathbf{s}_{N_1}^\mathrm{T} & -\hat{r}_{N_1}
\end{array}\right]\in \mathbb{R}^{N_1\times 3}, \\
\mathbf{z} &=[\mathbf{p}^\mathrm{T}, \|\mathbf{p}\|]^\mathrm{T} \in \mathbb{R}^{3\times 1}, \\
\mathbf{e} &=\left[e_{1,} e_2, \ldots, e_{N_1}\right]^\mathrm{T} \in \mathbb{R}^{N_1\times 1},
\end{aligned}
\end{equation}
and
\begin{equation}
\mathbf{B}=\left\|\mathbf{p}\right\| \cdot \mathbf{I}_{N_1}-\operatorname{diag}\left(\left[\hat{r}_1, \hat{r}_2, \ldots, \hat{r}_{N_1}\right]\right) \in \mathbb{R}^{N_1\times N_1}. \label{Bmatr}
\end{equation}
Following the above steps, the target localization problem can be addressed by solving for \(\mathbf{z}\) that satisfies \eqref{asd}. However, \(\mathbf{p}\) and its magnitude \(\|\mathbf{p}\|\) are both coupled in the vector \(\mathbf{z}\), which significantly increases the computational complexity of solving \eqref{asd}. \revise{To circumvent the challenge arising from the intrinsic nonlinearity, we initially treat \(\mathbf{p}\) and \(\|\mathbf{p}\|\) as independent variables, yielding a closed-form coarse estimate of the target's location. Subsequently, this coarse estimate is refined by reinstating the coupling relationship between \(\mathbf{p}\) and \(\|\mathbf{p}\|\).}  Specifically, under the assumption that \(\mathbf{p}\) and \(\|\mathbf{p}\|\) are mutually independent, \(\mathbf{z}\) can be estimated using WLS approach, by
\begin{equation}\label{firstWLS}
    \min _{\bm{\mathbf{z}} \in \mathbb{R}^{3}}(\bm{\alpha}-\mathbf{A} \mathbf{z})^\mathrm{T} \mathbf{R}^{-1}(\bm{\alpha}-\mathbf{A} \mathbf{z}),
\end{equation}
where $\mathbf{Q}=\operatorname{diag}\left(\left[\sigma_1^2, \sigma_2^2, \ldots, \sigma_{N_1}^2\right]\right)$ and $\mathbf{R}\approx{\mathbf{B}} \mathbf{Q} {\mathbf{B}}^\mathrm{T}$. 
It is evident that the  $\hat{\mathbf{z}}$ can be obtained by setting the gradient with respect to $\mathbf{z}$ to zero, by
\begin{equation}\label{eqforhatz}
   \hat{\mathbf{z}} =\left(\mathbf{A}^\mathrm{T} \mathbf{R}^{-1} \mathbf{A}\right)^{-1} \mathbf{A}^\mathrm{T} \mathbf{R}^{-1} \bm{\alpha}.
\end{equation}
However, the matrix $\mathbf{R}$ incorporates $\left\|\mathbf{p}\right\|$ (included in $\mathbf{B}$), which is still unknown and thus complicates the direct estimation. To tackle this issue, one can employ a two-stage estimation (TSE) approach to determine $\hat{\mathbf{z}}$ \cite{10714398}. 
\revise{To start with, we compute $\hat{\mathbf{z}}$ via \eqref{eqforhatz} with an initiali $\mathbf{R}^{(0)}=\mathbf{Q}$, followed by determining the matrix $\mathbf{B}$ via \eqref{Bmatr}. Subsequently, the covariance matrix $\mathbf{R}$ is updated via $\mathbf{R}\approx\mathbf{B} \mathbf{Q} \mathbf{B}^\mathrm{T}$. 
This iteration is terminated unitil $\|\mathbf{z}^{(\ell+1)} - \mathbf{z}^{(\ell)}\|_2 \leq \epsilon$ is satisfied with $\ell$ and $\epsilon$ being the iteration index and tolerance threshold, respectively.} 

Next, we refine the estimations by taking into account that $\mathbf{p}$ and $\|\mathbf{p}\|$ are coupled. We further express the estimation $\mathbf{\hat{z}}$ as   
\begin{equation}\label{equhatz}
    \hat{\mathbf{z}}=\left[\begin{array}{l}
\hat{x} \\
\hat{y} \\
\hat{v}
\end{array}\right]=\left[\begin{array}{l}
x+n_1 \\
y+n_2 \\
v+n_3
\end{array}\right],
\end{equation}
in which $\hat{x}$, $\hat{y}$, and $\hat{v}$ are the estimations corresponding to their counterparts with $\mathbf{n} = [n_1 , n_2 ,n_3]^\mathrm{T}$ being the Gaussian estimation errors.
Then, we can construct a vector $\mathbf{g}$, given by 
\begin{equation}\label{equgerror}
\mathbf{g}=\hat{\boldsymbol{\theta}}-\mathbf{S} \boldsymbol{ \upsilon},
\end{equation}
where $\hat{\boldsymbol{\theta}}=\left[\hat{x}^2, \hat{y}^2, \hat{v}^2\right]^\mathrm{T}, \boldsymbol{ \upsilon}=\left[x^2, y^2\right]^\mathrm{T}$ and
$$
\mathbf{S}=\left[\begin{array}{ll}
1 & 0 \\
0 & 1 \\
1 & 1
\end{array}\right].
$$
Substituting \eqref{equhatz} to  \eqref{equgerror} and omitting the second-order terms of error, $\mathbf{g}$ can be approximated as 
\begin{equation}\label{secondWLS}
\mathbf{g}=\left[\begin{array}{l}
2 x n_1+n_1^2 \\
2 y n_2+n_2^2 \\
2 v n_3+n_3^2
\end{array}\right] \approx\left[\begin{array}{l}
2 x n_1 \\
2 y n_2 \\
2 v n_3
\end{array}\right].
\end{equation}
Our goal is to estimate $\boldsymbol{ \upsilon}$  containing the true value of $x$ and $y$, which can be addressed by solving another WLS problem 
\begin{equation}\label{sectWLS}
    \min_{\boldsymbol{ \upsilon} \in \mathbb{R}^{2}}({\boldsymbol{\hat{\theta}}}-\mathbf{S} \boldsymbol{ \upsilon})^\mathrm{T} \mathbf{\Omega}^{-1}({\boldsymbol{\hat{\theta}}}-\mathbf{S}\boldsymbol{ \upsilon}).
\end{equation}
Similar to problem \eqref{firstWLS}, the estimate $\boldsymbol{\hat{\upsilon}}$ of problem \eqref{sectWLS} can be straightforwardly expressed as 
\begin{equation}\label{equgerrorsolve}
\hat{\boldsymbol{\upsilon}}=\left(\mathbf{S}^\mathrm{T} \mathbf{\Omega}^{-1} \mathbf{S}\right)^{-1} \mathbf{S}^\mathrm{T} \mathbf{\Omega}^{-1} \hat{\boldsymbol{\theta}},
\end{equation}
where \begin{align}
    \boldsymbol{\Omega}=\mathbb{E}\left(\mathbf{g g}^\mathrm{T}\right) \approx 4 \bar{\mathbf{C}} \operatorname{cov}(\mathbf{\hat{z}}) \bar{\mathbf{C}}. \label{omega}
\end{align} In \eqref{omega}, $\bar{\mathbf{C}}=\operatorname{diag}([\hat{x}, 
\hat{y}, \hat{v}])$, and $\operatorname{cov}(\mathbf{\hat{z}}) $ denotes  the covariance of $\hat{\mathbf{z}}$, written as 
\begin{equation}
    \operatorname{cov}(\hat{\mathbf{z}})=\mathbb{E}\left(\mathbf{n n}^\mathrm{T}\right) \approx\left({{\mathbf{A}}}^\mathrm{T} {\mathbf{R}}^{-1} {\mathbf{A}}\right)^{-1}.
\end{equation} As a result, the target position estimate $\hat{\mathbf{p}}$ is obtained by 
\begin{equation}\label{phat1}
    \hat{\mathbf{p}}=[\hat{x}, \hat{y}]=\left[\pm \sqrt{\hat{\boldsymbol{\upsilon}}_1}, \pm \sqrt{\hat{\boldsymbol{\upsilon}}_2}\right],
\end{equation}
where ${\hat{\boldsymbol{\upsilon}}_i}$ ($i=1,2$) is the $i$-th element of $\hat{\boldsymbol{\upsilon}}$.
We can determine the correct value by evaluating their accumulative squared error, which is 
\begin{equation}  \label{phat2}      
\mathcal{L}=\sum_{i=1}^{N_1}(\left\|\hat{\mathbf{p}}-\mathbf{s}_i\right\|+\left\|\hat{\mathbf{p}}\right\|-\hat{r}_i)^2.
\end{equation}
The ultimate estimation of the target's position corresponds to the value of $\hat{\mathbf{p}}$ that minimizes the function $\mathcal{L}$.

Building upon the above, we estimate the Tx velocity based on geometric relationships and the Doppler shifts associated with the LoS and NLoS. 
Combining \eqref{eqforridot}, \eqref{eqfordidot},  and \eqref{eqforerror}, we have 
 \begin{equation}\label{eqlforV}
     \mathbf{{{u}}}_i = \mathbf{C}_i\mathbf{\dot{s}}_i-\mathbf{{e}'}_i,
 \end{equation}
where $   \mathbf{u}_i =\left[\hat{\dot{r}}_{i}, \hat{\dot{d}}_{i}\right]^\mathrm{T}$, $
\mathbf{{e}'}_i =\left[e_{\dot{r}_i}, e_{\dot{d}_i}\right]^\mathrm{T}$ and $\mathbf{C}_i$ is 
 \begin{equation}
\begin{aligned}
\mathbf{C}_i &=\left[\begin{array}{c}
\boldsymbol{\rho}^\mathrm{T}_{1,_i};
\boldsymbol{\rho}^\mathrm{T}_{2,_i} 
\end{array}\right]=\left[\begin{array}{cc}
     \frac{(\mathbf{s}_i-\mathbf{\hat{p}})^\mathrm{T}}{\left\|\mathbf{s}_i-\mathbf{\hat{p}}\right\|};
    \frac{(\mathbf{s}_i-\mathbf{t})^\mathrm{T}}{\left\|\mathbf{s}_i-\mathbf{t}\right\|} 
\end{array} \right].
\end{aligned}
\end{equation}
The goal is to solve $\mathbf{\dot{s}}_i$ from \eqref{eqlforV}, which can be regarded as an LS estimation problem. Hence, the solution can be straightforwardly articulated as
 \begin{equation}\label{sfinalhat}
     {\hat{{\dot{\mathbf{s}}}}}_i = (\mathbf{C}_i^\mathrm{T}\mathbf{C}_i)^{-1}\mathbf{C}_i^\mathrm{T}\mathbf{u}_i.
 \end{equation}
To sum up, the overall procedure of the target position and mobile Tx velocity estimation algorithm is presented in Algorithm \ref{alg3}.

\begin{figure}[t]
  \begin{minipage}{\linewidth}
  \begin{algorithm}[H]
\renewcommand{\algorithmicrequire}{\textbf{Input:}}
    \renewcommand{\algorithmicensure}{\textbf{Output:}}
    \caption{Target positioning and Tx velocity estimation method} 
    \label{alg3} 
    \begin{algorithmic}[1]
      \REQUIRE $\hat{\mathbf{r}}$ (derived from delay parameters); $\hat{\dot{r}}_{i}$ and $\hat{\dot{d}}_{i}$ (derived from Doppler parameters), $\mathbf{R}^{(0)}$
      \REPEAT
      \STATE Based on \eqref{eqforhatz} and $\mathbf{R}$ to obtain a rough estimate $\hat{\mathbf{z}}$ about the target position. 
      \STATE Update $\mathbf{B}$ and $\mathbf{R}$ via \eqref{Bmatr} and $\mathbf{R}\approx\mathbf{B} \mathbf{Q} \mathbf{B}^\mathrm{T}$.
      \UNTIL{Converged}
       \STATE Refine the position estimation via \eqref{phat1} and check the correct value via  \eqref{phat2}.
      \STATE Calculate the Tx velocity $\hat{{\dot{\mathbf{s}}}}$ according to \eqref{sfinalhat}
      \ENSURE Target position $\hat{\mathbf{p}}$ and Tx velocity $\hat{{\dot{\mathbf{s}}}}$.
    \end{algorithmic} 
  \end{algorithm}
  \end{minipage}
\end{figure}

\section{Experiment Results}
In this section, we evaluate the proposed environment sensing design through extensive experiments. Specifically, we begin with a concise description of the experimental hardware setup, followed by demonstrating experimental results.
\subsection{Experimental Hardware Setup}
\begin{figure}[t]
	\centering
	\includegraphics[width=0.9\columnwidth]{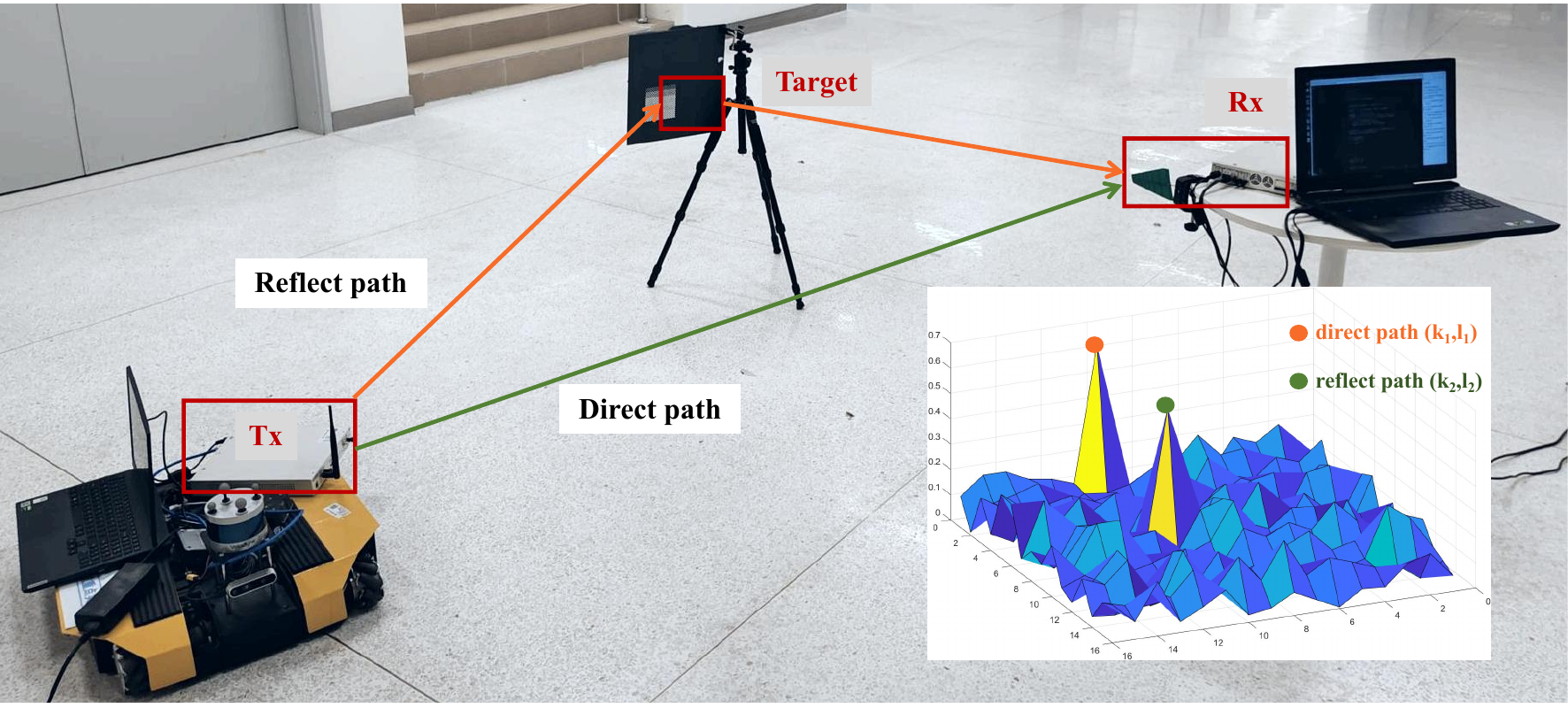}
	\caption{The environment sensing physical scene built upon the SDR platform with USRP device using OTFS signals.}
	\label{USRP}
\end{figure}

 The experimental setup is illustrated in Fig. \ref{USRP}, where a robotic vehicle equipped with USRP hardware serves as the Tx, and the Rx is implemented using another USRP device. The target object, depicted as a metal plate, is placed on a tripod to simulate a reflective surface. The setup highlights both the direct signal path between Tx and Rx and the reflected path via the target, providing a practical environment to evaluate the performance of the proposed sensing design.

\begin{table}[t]
\renewcommand{\arraystretch}{1.3}
   \caption{}
   \caption*{DEFAULT SIMULATION PARAMETERS}
\label{table1}
\centering
\begin{tabular}{c|c}
  \hline \textbf{Parameters} & \textbf{Values} \\
  \hline     Number of time slots ($N$) & 1024 \\
 \hline  Number of subcarriers ($M$) & 1024 \\ 
\hline        Carrier frequency ($f_c$) & 5.6 GHz \\ \hline      Subcarrier space ($\Delta f$) & 93.75 KHz \\
 \hline       Signal bandwidth ($B$) & 96 MHz \\
  Doppler resolution ($\Delta \upsilon $) & 91.55 Hz \\
 \hline  Delay resolution ($\Delta \tau$) & 10.42 ns \\
 \hline
\end{tabular}
\label{CMP}
\end{table}

\begin{figure}[t]
	\centering
	\includegraphics[width=1.1\columnwidth]{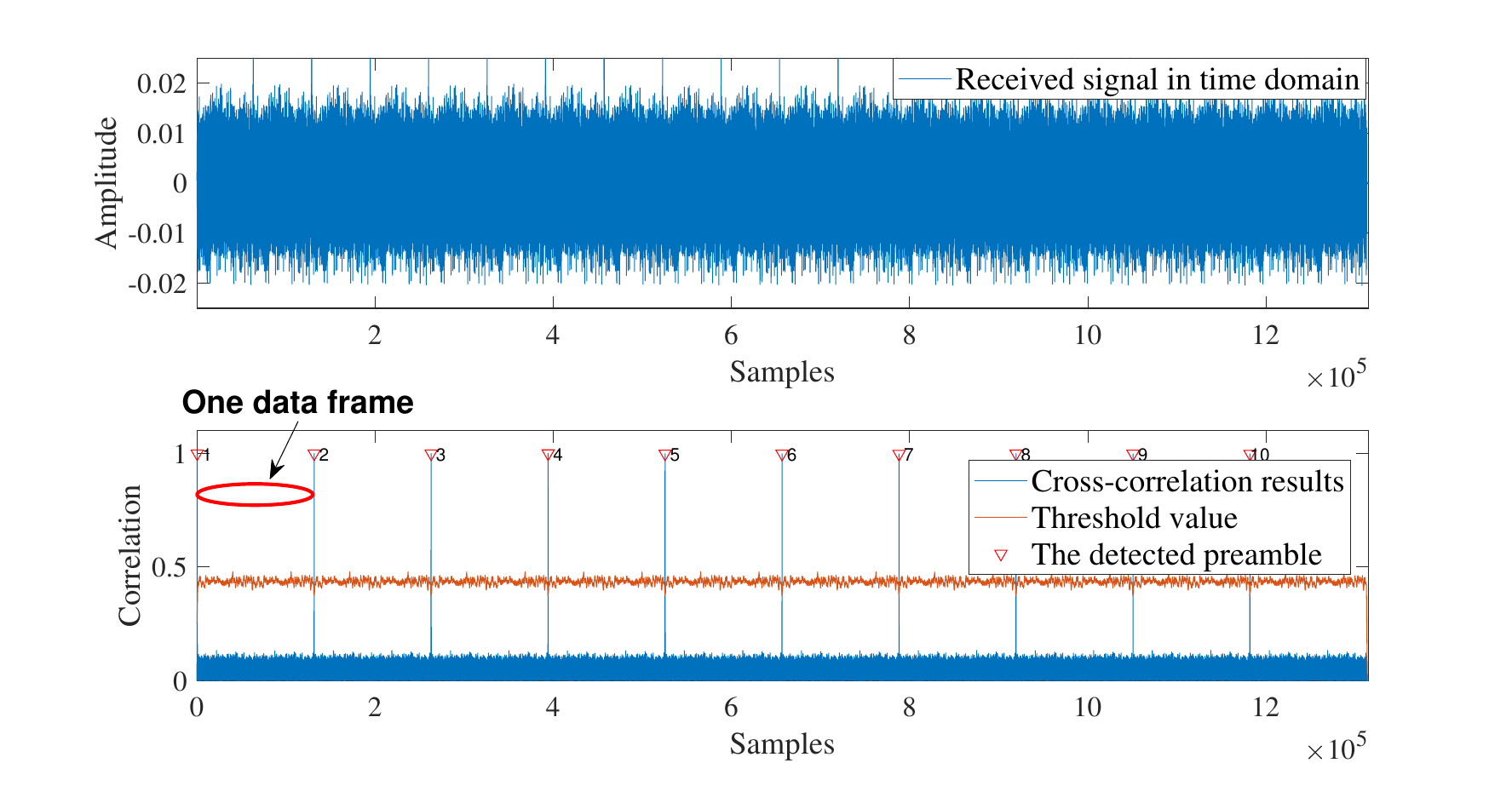}
	\caption{The received time-domain signals via  USRP and the corresponding cross-correlation results.}
	\label{timeDomainyn}
\end{figure}

We employ the National Instruments SDR platform with USRP-2954 devices to implement the OTFS signal-based environment sensing framework. The
USRP-2954 supports RF frequencies ranging from $10$ MHz to $6$ GHz with a maximum output power of $17$ dBm (50 mW). This platform enables the emulation of diverse modulation schemes, facilitating adaptability to various communication scenarios. In this setup, a full-phase antenna is utilized at the Tx, while a directional antenna is deployed at the Rx to enhance the quality and reliability of the collected signals. Unless otherwise specified, experimental parameters are shown in TABLE \ref{CMP}. The maximum speed of the mobile Tx is set to $260$ km/h. \revise{ The theoretical distance resolution $d_0$ and the velocity component resolution $v_0$ are obtained by $d_0=\Delta \tau c=3.12$ m and $v_0=\frac{\Delta \upsilon c}{f_c}=4.9$ m/s, respectively, in which the delay resolution $  \Delta\tau = \frac{1}{1024 \times 93.75 \times 10^3} \approx 10.42\ \text{ns}$ and  the Doppler resolution $ \Delta\nu = \frac{93.75 \times 10^3}{1024} \approx 91.55\ \text{Hz}$.} The distance interval between Tx and Rx is set to range from $10$ m to $80$ m and the target is not allowed to be in the line from Tx to Rx. The size of the pilot guard space is set to $32\times 32$. 
The receive gain is fixed at $0$ dBm, and the maximum available power budget at the Tx is $50$ mW\cite{10118889}. The received pilot signal at the Rx is shown in Fig. \ref{USRP}, which contains two significant channel responses while affected by noise. In most cases, the amplitude of the pilot response is much greater than that of the noise, ensuring reliable channel estimation.
Additionally, the Tx power ranges from 3 dBm to 17 dBm.

 After data acquisition, the received signals are processed in MATLAB 2021 to estimate channel parameters in the DD domain, followed by performing target localization. The time-domain signals captured from the USRP are depicted in the top panel of Fig. \ref{timeDomainyn}, comprising approximately \(1.31 \times 10^6\) sampling points. Note that, accurate parameter estimation requires precise identification and separation of each data frame. To this end, a preamble signal is embedded at the beginning of each frame at the Tx end. At the Rx end, the preamble signal is detected by computing the cross-correlation between the received time-domain signal and the known preamble, as shown in the bottom panel of Fig. \ref{timeDomainyn}. To ensure robust preamble detection, a constant false alarm rate (CFAR) of \(10^{-5}\) is applied, with the corresponding detection threshold demonstrated as the orange curve in the bottom panel of Fig. \ref{timeDomainyn}. The locations of the preamble signals are identified as the cross-correlation peaks exceeding the threshold, thereby enabling accurate frame synchronization and effective environment sensing in the DD domain.

\subsection{Environment Sensing Performance}

\begin{figure}[tbp]
	\centering
	\includegraphics[width=\columnwidth]{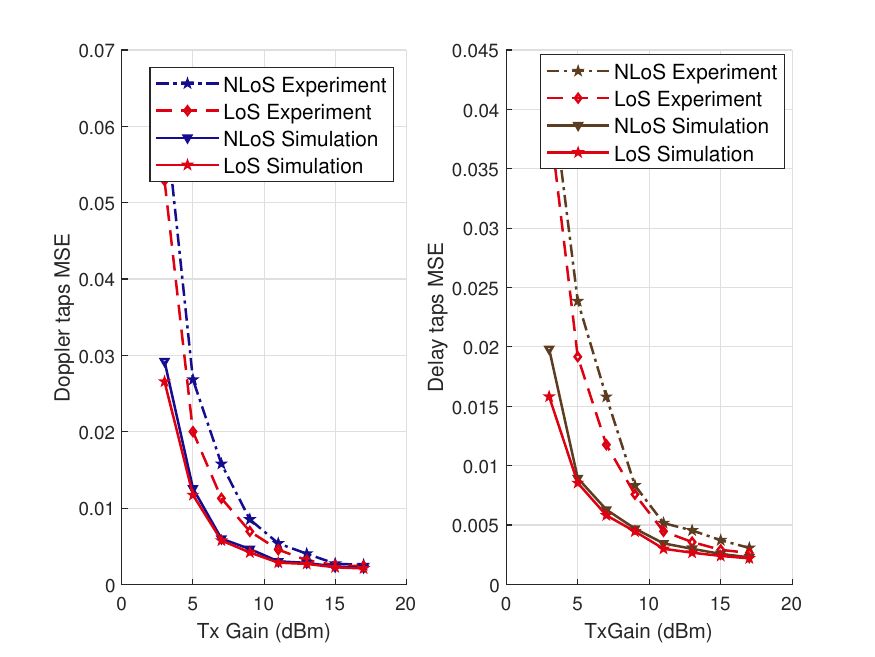}
	\caption{The MSE performance of the estimated delay and Doppler taps.}
	\label{MSE_kl_log}
\end{figure}


Fig. \ref{MSE_kl_log} evaluates the proposed environment sensing design using mean square error (MSE) of the estimated delay and Doppler shifts as performance metrics. We demonstrate the estimation results derived from ideal simulated data and experimental data collected using the SDR-USRP platform. As anticipated, the estimation performance based on simulated data I superior to that of the experimental data due to the interference and noise inherent in real-world environments. Another observation is that the gap between the simulation and experimental results narrows as the transmission power increases. This is attributed to the enhanced signal-to-noise ratio (SNR) at higher transmit power levels, which mitigates the effects of interference and noise. Furthermore, the estimation performance of the LoS path consistently outperforms that of the NLoS path, both in simulations and experiments. The reason is that the LoS path experiences less signal attenuation and exhibits a simpler propagation model, with fewer multipath effects and lower levels of phase distortion. In summary, Fig. \ref{MSE_kl_log} validates the practical applicability of the proposed channel estimation algorithm, confirming its suitability for deployment in real-world applications.

\begin{figure}[tbp]
	\centering
	\includegraphics[width=0.9\columnwidth]{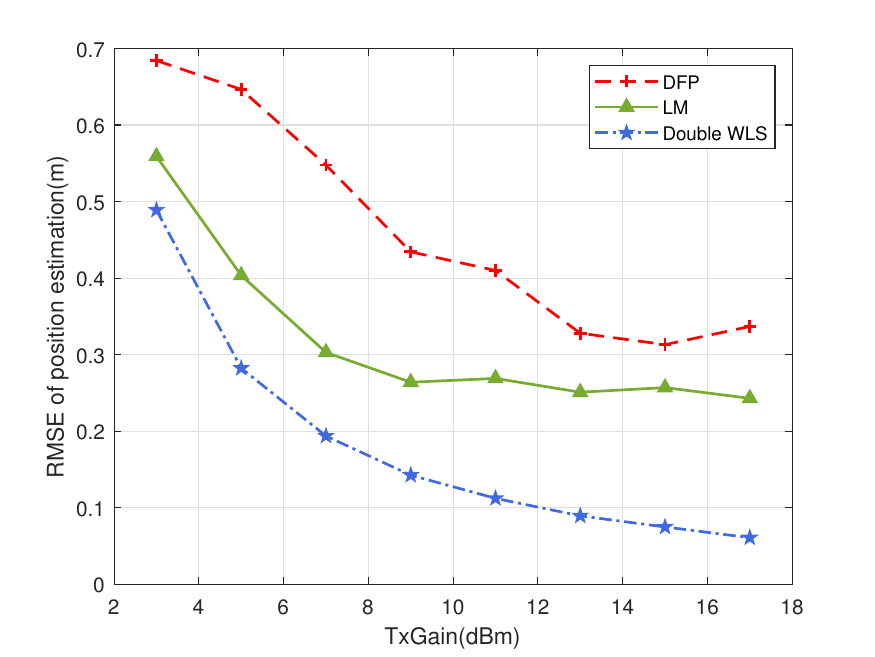}
	\caption{The RMSE of target location estimation versus different transmit powers with various algorithms.}
	\label{Error_P}
\end{figure}

\begin{figure}[t]
	\centering
	\includegraphics[width=0.9\columnwidth]{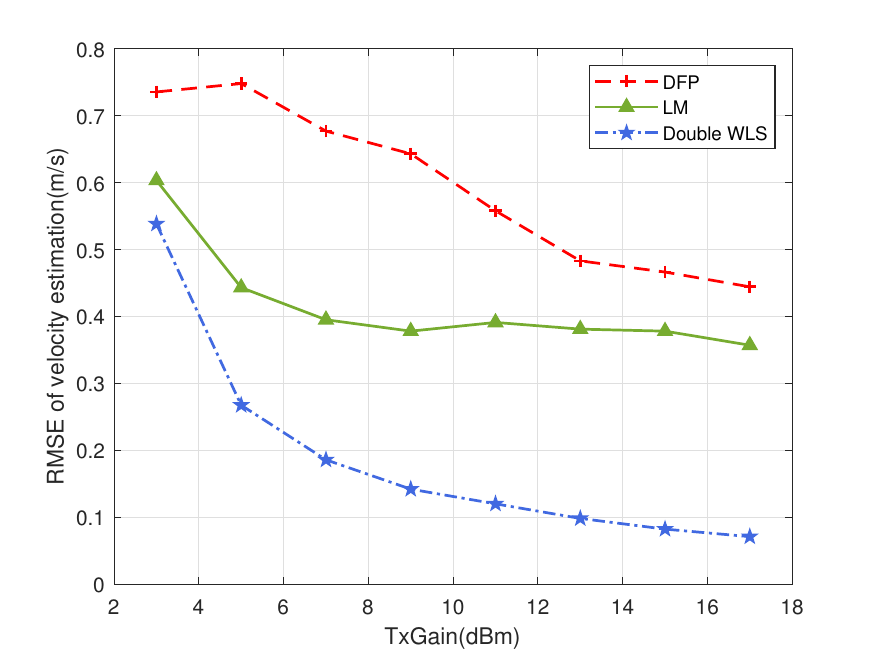}
	\caption{The RMSE of velocity estimation versus transmit powers with different location determining algorithms.}
	\label{Error_S}
\end{figure}

\begin{figure*}[tbp]
    \centering
    \subfloat[\label{cosRelationkl}]{\includegraphics[width=0.35\textwidth]{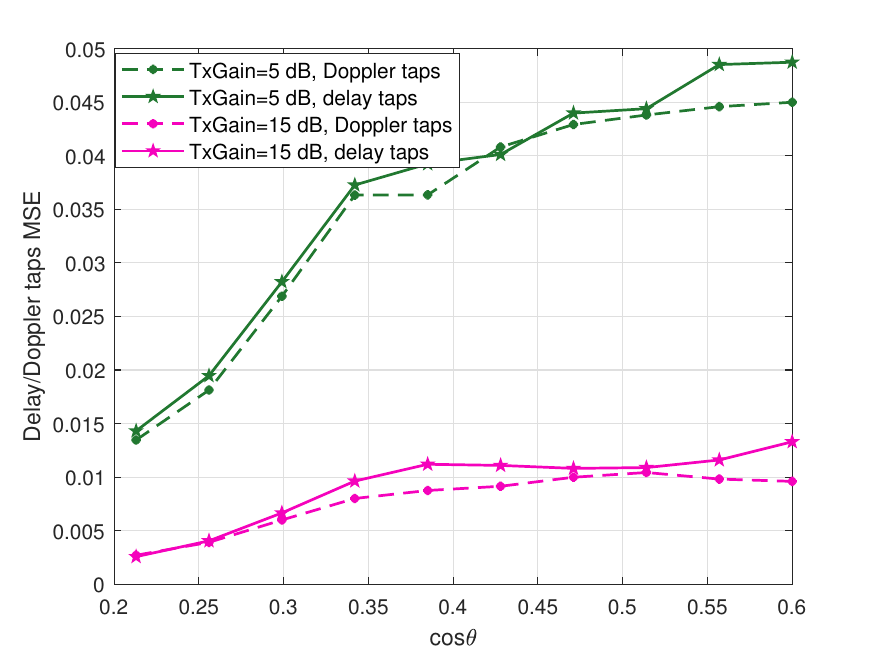}}
    \subfloat[\label{RMSE_P_cos}]{\includegraphics[width=0.35\textwidth]{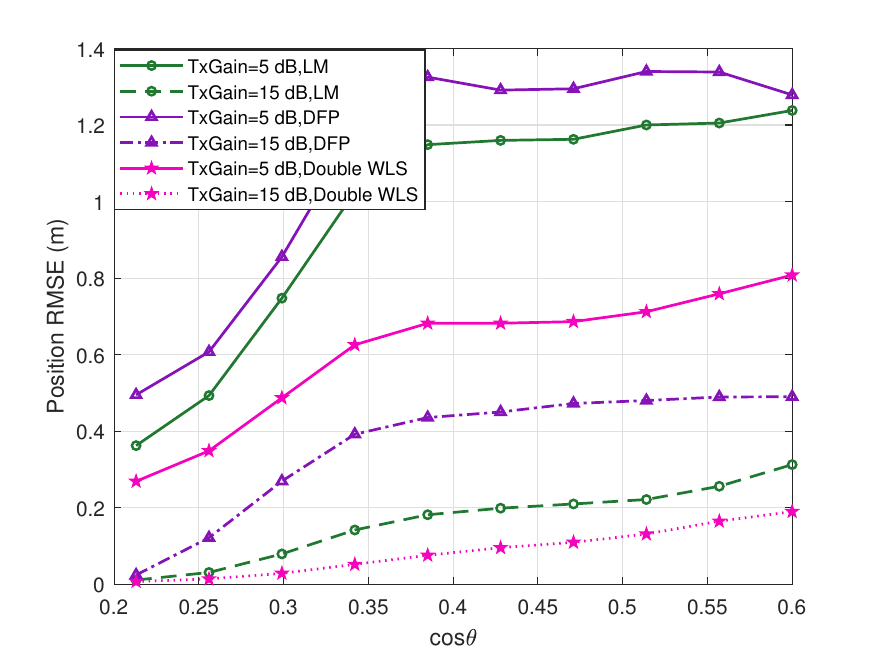}}
    \subfloat[\label{RMSE_V_cos}]{\includegraphics[width=0.35\textwidth]{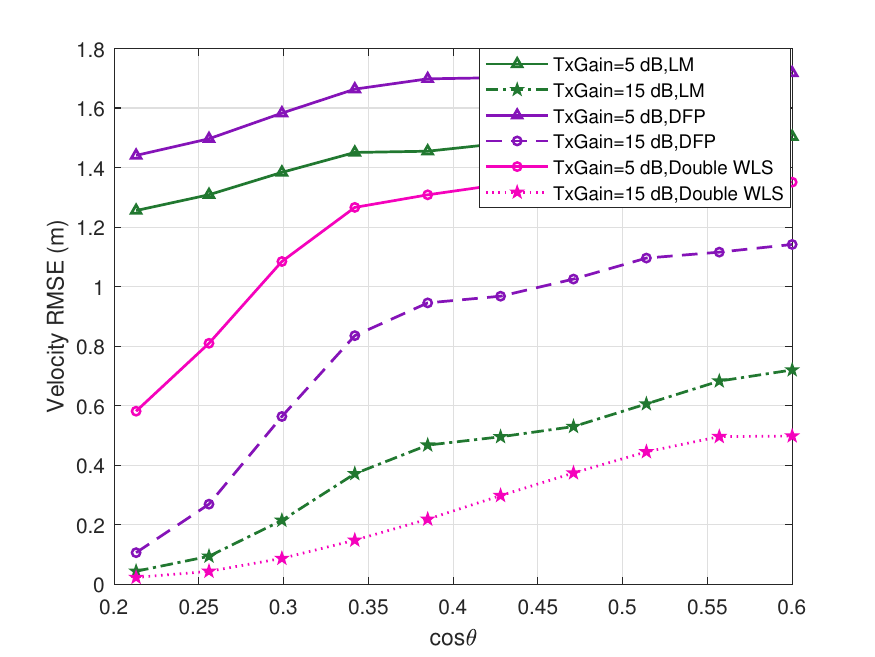}}
    
    \caption{(a). The MSE of the estimated delay and Doppler taps versus the values of cos$\theta$ with different transmit powers. (b). The RMSE of position estimation against different values of cos$\theta$  under varying transmit powers. (c). The RMSE of velocity estimation versus different values of cos$\theta$ under various Tx powers.}
    \label{cosRelation}
\end{figure*}

Then, we examine the environment sensing performance by leveraging the estimated delay and Doppler shifts based on the collected data by USRP. In particular, we compare the proposed double WLS with two benchmark methods, i.e., the Levenberg-Marquardt (LM) algorithm \cite{amini2018efficient} and the Davidon-Fletcher-Powell (DFP) algorithm \cite{mamat2018derivative}, which are employed to solve equations \eqref{asd} respectively. The LM algorithm is known for its robustness in solving nonlinear least squares problems, while the DFP algorithm, a quasi-Newton method, approximates the Hessian matrix efficiently to facilitate optimization for smooth objective functions.

In Fig. \ref{Error_P}, we present the position estimation performance in terms of root mean square error (RMSE) across varying transmit power levels. The results demonstrate that all evaluated methods obtain enhanced performance with increasing transmit power. In particular, the proposed double WLS approach consistently achieves the highest estimation accuracies across all power levels. This is attributed to its ability to integrate both coarse and refined parameter estimations, thereby effectively mitigating residual errors. In contrast, the DFP algorithm exhibits the worst performance due to its sensitivity to noise and suboptimal convergence behavior. Another interesting finding is that the DFP algorithm converges to an error floor with an RMSE of approximately $0.25$ m when the transmit power exceeds $9$ dBm. This is because the DFP algorithm relies on approximate gradient and Hessian information, degrading its robustness in solving non-linear optimization problems. In contrast, our double WLS method continues to improve with the power increasing, ultimately achieving an RMSE below $0.1$ m. 

In Fig. \ref{Error_S}, we show the RMSE results of Tx velocity estimation based on the estimated position $\mathbf{\hat{p}}$ obtained from the LM, DFP, and the proposed double WLS approach.
 As illustrated in the figure, the RMSE trends of velocity estimation closely match those of position estimation, indicating that position accuracy contributes directly to velocity estimation precision. Among the evaluated methods, the proposed algorithm consistently outperforms the alternatives, achieving the lowest RMSE for velocity estimation. Notably, as the transmit power exceeds 15 dBm, the RMSE of the proposed method is lower than 0.1 m/s, in contrast to approximately 0.45 m/s for the DFP algorithm and around 0.38 m/s for the LM method. From  Fig. \ref{Error_P} and Fig. \ref{Error_S}, it is safe to conclude that the proposed positioning algorithm can effectively perform sensing functionality on the SDR-USRP platform.


We further validate the impact of data characteristics on algorithm performance, particularly induced by the variations in the target and Tx positions. Towards this end, we define the angle \(\theta\) to characterize the coupling between the LoS and NLoS paths at the Tx, which is
\[
\theta = \arccos\frac{\|\mathbf{s}-\mathbf{t}\|^2 + \|\mathbf{s}-\mathbf{p}\|^2 - \|\mathbf{p}-\mathbf{t}\|^2}{2\|\mathbf{s}-\mathbf{p}\| \|\mathbf{s}-\mathbf{t}\|}, \quad \theta \in [0, \frac{\pi}{2}].
\]
Intuitively, when \(\theta = 0\), the target lies directly along the line from the Tx to the Rx, overlapping the LoS and NLoS paths. In such cases, the received signal at the Rx may fail to reveal the target’s physical characteristics.
Furthermore, for small values of \(\theta\), the delay and Doppler shift parameters of the LoS and NLoS paths become nearly identical. Under this situation, the response peaks of the two paths are closely located to each other, complicating delay and Doppler taps estimation and degrading the environment sensing performance.

To verify this, we show the MSE of delay and Doppler taps estimations against various \(\cos \theta\) values in Fig. \ref{cosRelationkl}. The results demonstrate that all methods experience performance degradation as \(\cos \theta\) increases, which aligns with our earlier discussion. This occurs as \(\cos \theta \rightarrow 1\) when \(\theta \rightarrow 0^\circ\), resulting in stronger coupling between the LoS and NLoS paths. The increased coupling reduces the separability of the two paths, thereby degrading the estimation accuracies of both delay and Doppler taps. Moreover, we observe that the influence of $\cos \theta$ can be mitigated at high SNRs, which improves the system's ability to distinguish between the LoS and NLoS components, even under strong coupling conditions.  

Next, in Fig. \ref{RMSE_P_cos} and Fig. \ref{RMSE_V_cos}, we present the RMSE for position and velocity estimation comparison between the LM algorithm and the proposed double WLS method. The trends for position and velocity estimation with increasing \(\cos\theta\) are similar to those for delay and Doppler tap estimation. Additionally, the RMSE results for both methods enhance significantly with higher transmit power, emphasizing the critical role of signal strength in improving estimation accuracy. Notably, the proposed double WLS method is superior to the LM algorithm, achieving substantially lower RMSE values across all transmit power levels and \(\cos\theta\) ranges, highlighting the reliability of our proposed double WLS method for position and velocity estimation.


\section{Conclusion}
This paper investigated OTFS-assisted environment sensing, focusing on channel estimation, target localization, and the Tx velocity determination. 
To this end, a pilot-based off-grid channel estimation approach was first explored to determine the delay and Doppler taps in the presence of fractional orders. Based on the estimated channel parameters, the target position is localized through the developed three-ellipse-based positioning algorithm, where a double WLS method is introduced to iteratively solve the associated nonlinear equation. The Tx velocity was subsequently estimated using the LS approach. To verify the effectiveness of the proposed design, an experimental system was established using the SDR-USRP platform. Extensive experimental results demonstrated that our proposed approach achieves precise delay and Doppler shift estimations and determines accurate target location and the Tx velocity, validating its potential in environment sensing applications. 

\ifCLASSOPTIONcaptionsoff
  \newpage
\fi

\bibliographystyle{IEEEtran}
\bibliography{ref}

\end{document}